Enhanced photoinduced nematic reorientation in mixture with azo-dye-substituted polymer.

A. Parfenov [a] *, A. Chrzanowska

*Diagnostic Instrumentation and Analysis Laboratory, 205 Research Blvd, Starkville, MS 39759, USA;

Krakow University of Technology Institute of Physics, Podchorazych 1, Krakow 30-084, Poland

The optically induced reorientation of nematic liquid crystal doped with azo-dye substituted polymer is investigated measuring photoinduced birefringence. These measurement reveal the value ($\Delta n \sim 0.1$) of induced birefringence of liquid crystal with dye polymer which significantly exceeds the value of birefringence previously obtained in nematic mixture with the low molecular weight dye. More than one order of enhancement is connected with lower diffusivity of polymer.

---

a) on leave from Lebedev Physics Institute, Russia, electronic mail: parfenov@neur.lpi.msk.su

The great interest in light-induced reorientation of nematic liquid crystals [1] is closely connected with large potential of this effect in such applications as nonlinear optics, photonics and optical data processing [2]. Optical reorientation is greatly enhanced when the liquid crystal is doped with a small value of light -absorbing dye [2-4]. It is also observed that this specific effect is strongly sensitive to the molecular structure of the guest-host, and mainly to dye structure [5]. It is established that diffusion of dye molecules in liquid crystal host has great influence on the observed reaction [6,7]. Recently the molecular theory [7] of the resonant and viscous torques in *trans-cis* systems with nematic liquid crystals has explained the main dependencies in terms of diffusional motion and hydrodynamic reorientation of molecules in a three component mixture (*trans-*, *cis-* conformers and nematic liquid crystal molecules).

We report the enhanced optically induced reorientation in liquid crystal doped with dye - substituted polymer and corresponding induced birefringence. The experiments are performed with a homeotropic LC cell, containing nematic 5CB from Merck (birefringence $\Delta n=0.2$) and azobenzene substituted polymer poly-Orange Tom-1 Isophoronedisocyanate. The LC - polymer mixture contains 3-5 –wt.% of the latter substance. The molar part of azobenzene in this polymer is estimated as ca 50% in weight. The spacers fixing the thickness of this cell L are of 34 μm. Induced birefringence is observed in a polarised probe beam of a diode laser. The absorption spectra for this polymers presented in [8] as verified coincides with the spectra of its solution in the LC (maximum absorption density achieved 1.8-2.0 in the solution). Wavelength of laser diode radiation (630 nm) is far from absorption band of polymer solution. The optical response of the media is measured under exposure by linearly polarised light of the wavelength 488 nm located within the absorption band of the azobenzene group. The diameter of the illuminated area and light intensity are of 4-9 mm and of 2-7 mW/cm$^2$ respectively. The polarization plane of diode laser radiation is set at 45 degrees to the green beam polarization plane (polarization planes are denoted in Fig.1 by arrows).

The LC cell transmission for the λ=630 nm light is observed between polarisers. The change of the transmission is attributed to the birefringence arising due to the declination of nematic LC from the homeotropic alignment at the angle θ. This angle determines the induced birefringence as follow [9]:

$$\Delta n(\theta) = (\cos^2\theta/n_o^2 + \sin^2\theta/n_e^2)^{-1/2} - n_o, \qquad (1)$$

with $n_o$ and $n_e$, refractive indices for ordinary and extraordinary optical beams in liquid crystal media.

The cell transmission in birefringence regime is determined by induced phase shift $\Delta\Phi=2\pi \Delta n(\theta)L/\lambda$ between two polarization components (chosen at 45 degrees to the polarizer axis):

$$I=I_0 T_0 \sin^2(\Delta\Phi/2)+I_{scat}, \qquad (2)$$

with $I_0$, intensity of incident light; $T_0$, transparency of polarizer for light polarized at 0 degrees; $I_{scat}$, light intensity scattered and partially depolarized by LC cell.

The change of the LC cell transmission (Equation 2) is measured in a simple scheme (optical scheme of the experiment is presented in Fig.1), which included the LC cell itself, green Ar laser (488 nm) for excitation and red diode laser (630 nm). A photodiode registers the transmitted power of the red beam. The signal from the photodiode is recorded on a plotter with swiping speed 1-6 cm/min, well suitable for realized response time. The excitation beam is incident almost normally, red readout beam - at a small angle of 5 degrees.

The induced birefringence $\Delta n(\theta)L$ is determined from induced phase shift $\pi \cdot \Delta n(\theta)L /\lambda \sim k\pi$, where k is a number of the specific oscillations (shown in the inset in Fig.1) appeared at ascend and decay of the optical response activated by a pulse of green light. The rise and decay of the induced phase shift can be well fitted by exponential dependencies of a time (Fig.2). The maximum value of induced birefringence $\Delta n(\theta)$ is estimated as 0.08-0.10 for light intensity 5 mW/cm$^2$. This value corresponds to declination from the initial state at the angel $\theta \sim 45$ degrees (determined by means of Equation 1 as a value averaged over the LC layer thickness). The LC texture created under the uniform illumination is not quite flat. Considerable part of the light power (up to 30% in some cases) is scattered; thereby diminishing the contrast of mentioned oscillations. An aggregation of polymer in a solution could be a reason for irregularity leading to this scattering.

Known results [10] are giving much lower induced birefringence ~0.005 for LC doped with azobenzene D2 dye in the comparable conditions (LC cell thickness ~ 37.5 μm, only 2-3x larger light intensity ~ 10 mW/cm$^2$). The rise/decay time published is of 1 sec.

The time constant of visco –elastic relaxation [9] of LC is $\tau \sim \gamma L^2 / K$, with $\gamma$, viscosity; $K$, elastic coefficient of LC and $L$ - its thickness. With the given values of thickness and material constants it is of order of 1 second as in mentioned experiments with monomers [10]. In our experiments the process of reorientation is characterized by the rise/decay times of 40 seconds and thus cannot be explained exclusively by visco-elastic relaxation. Dependencies of rise/decay time on the light intensity are presented in Fig.3 as well as the maximum phase shift achievable for given light intensity determined from specific oscillations at the optical response as mentioned above. The polymer dye in LC imposes longer excitation relaxation than in a case of monomer dye additive (40 sec compared to 1 sec in [10]), but it is much shorter than for light induced reorientation in azobenzene-containing polymer liquid crystals [11]. The measurements in the lower intensity region have obstacles in slowing of the LC - mixture response, while at higher intensities (5-6 mW/cm$^2$) the visibility of oscillations at the front of response becomes inferior.

The observed difference explanation is probably concluded in a role of polymer binding the dye molecules. According to the theory [7] the resonant contribution in optical torque is strongly dependent on the diffusion coefficients of *trans* and *cis* conformers of dye molecules (and their difference). At the same time the doping dye molecules can be introduced in the LC material in different forms- as a stand-alone monomer or in the polymer compound as in our experiments. In the last case the dye group are included as substitute side-groups. These groups being comparatively weakly bound with the main polymer chain retain much of the optical properties of genuine dye molecules, including absorption strength of oscillator for resonant wavelength. The azobenzene inserted in LC in the same molar portion as a monomer and as a polymer substitute acts optically in the same way (absorbs the light and perpetuates conformal transitions while the light is on). At the same time the diffusion coefficient for polymer dye must be significantly lower than for monomer. Actually the rotational diffusion is efficiently dumped due to interaction between neighboring side groups, while translation one is practically excluded for dye molecule attached to the polymer main chain. The dumping of rotational diffusion must strongly depend on the distance between side-groups attached to main polymer chain. This dependence can be tested directly if different homologues of the same polymer with different separation of substitute become available.

The lower diffusion coefficient for dye -substituted polymer (compared to the monomers) leads to a lesser migration of dye-initiated excitations and as a result to a larger value of the resonant torque per a mixture volume in accordance with the mentioned theory. Knowledge of diffusion coefficients for polymer and corresponding dye monomer could provide better quantitative comparison. Unfortunately these coefficients are still to be measured for both polymer and monomer dissolved in LC.

Komitov [12] reports on the planar-homeotropic transition in the cell containing azobenzene-based LC with ability of photoconformation, when the adsorption of *cis*-conformers on the LC-substrate interface prevails over *trans*- conformer adsorption due to the larger dipole moment. We can not exclude similar behavior of the dye in our case, which could induce some stabilization of the homeotropic alignment. At the same time such an adsorption is originated from the interaction with the substrates. In the case of ITO coated substrates it is logical to assume that this interaction has the nature of electrostatic attraction to the conductive surface. As we are taking uncoated glasses this stabilizing effect must be effectively reduced. Also in LC-dye mixture the role of this surface adsorption should not be as large as for the cited case for the used concentrations of dye (ca 5 weight % and only small of them are statistically under illumination in *cis*-form, which causes the stabilization of homeotropic orientation). Thus to our opinion this stabilization effect should not contribute significantly in observed LC reorientation.

In conclusions the azo-dye polymer is used as a photoactive additive to the nematic LC. Under the resonant illumination this mixture exhibits induced birefringence. Comparison with the published results on the LC mixtures containing monomer azobenzene dyes shows significant gain in the value of induced birefringence. This observation of the enhanced induced birefringence in LC-polymer mixture qualitatively proves the theory [7]. Practically the sensitivity enhancement for LC-polymer mixture (e.g. the threshold of reorientation is only of 2 mW/cm$^2$) directly leads to applications in optical technology. For example lower light power requirements open an application for fiber optics switches and routers as well as optical image processing via nonlinear filtering [13] or dynamic holography.

The Association for Super-Advanced Electronics Technologies (ASET) and the New Energy Development Organization (NEDO) supported in part this study. We thank Electrotechnical Laboratory and Dr. H.Yokoyama for support.

Figures captions:

Figure 1. Optical scheme of the experiment: P –polarizer, PD – photodiode.

Figure 2. Phase shift changes under exposure for the selected intensity of light (5 mW/cm$^2$). Both rise and decay are well fitted by the exponential dependence with time constant 40 s.

Figure 3. Intensity dependencies of rise/decay times vs light intensity determined as a time constant $\tau$ with approximation by $\exp(-t/\tau)$ and maximum phase shift achieved at given illuminating light intensity.

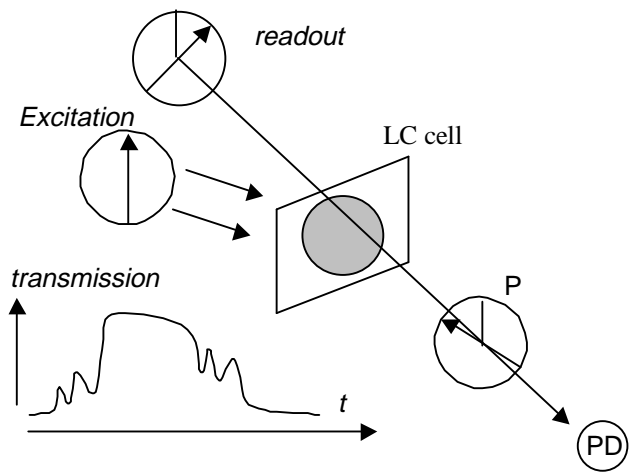

Fig.1. A.Parfenov, A.Chrzhanowska

Enhanced photoinduced …..

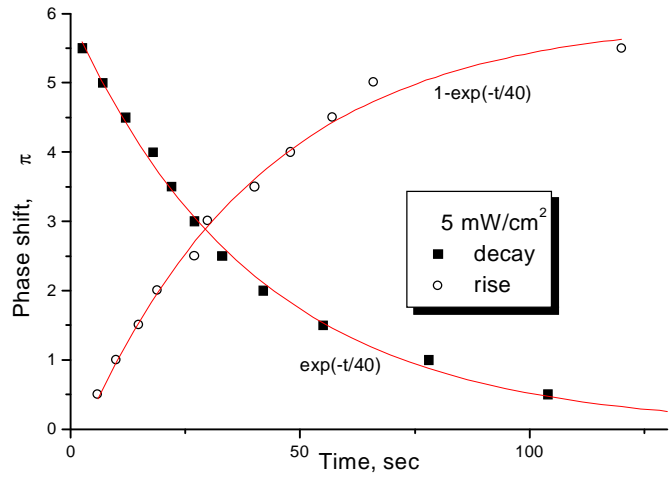

Fig.2. A.Parfenov, A.Chrzhanowska

Enhanced photoinduced …..

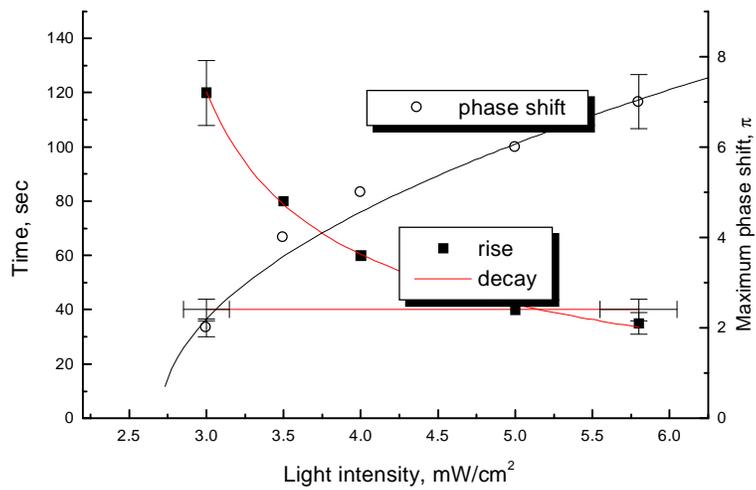

Fig.3. A.Parfenov, A.Chrzhanowska

Enhanced photoinduced …..